\documentclass[aps, prb, nofootinbib, twocolumn]{revtex4-2}
\usepackage[utf8x]{inputenc}
\usepackage{amsmath}
\usepackage{amsfonts}
\usepackage{textcomp}
\usepackage{amssymb}
\usepackage{empheq}
\usepackage{esvect}
\usepackage{tikz}
\usepackage{subcaption}
\usetikzlibrary{calc}
\usepackage{xcolor}
\definecolor{lightpink}{RGB}{255, 230, 230}
\usepackage{hyperref}
\hypersetup{
  colorlinks   = true, 
  urlcolor     = blue, 
  linkcolor    = blue, 
  citecolor   = red 
}
\usepackage{tabularx}
\usepackage{multirow, makecell}

\newcommand{\p}{\partial}
\newcommand{\f}{\frac}
\newcommand{\s}{\sqrt}

\newcommand{\be}{\beta}
\newcommand{\al}{\alpha}
\newcommand{\rot}{\operatorname{rot}}
\renewcommand{\div}{\operatorname{div}}
\newcommand{\grad}{\ve{\nabla}}

\newcommand{\pp}{\perp}
\renewcommand{\pl}{\parallel}

\renewcommand{\c}{\mathcal}
\newcommand{\ve}{\boldsymbol}

\newcommand{\D}{\Delta}

\title{Electromagnetic sources beyond common multipoles}

\begin{document}
\author{Nikita A. Nemkov}
\email{nnemkov@gmail.com}
\affiliation{Russian Quantum Center, Skolkovo, Moscow 143025, Russia}
\author{Vassili A. Fedotov}
\email{vaf@orc.soton.ac.uk}
\affiliation{Optoelectronics Research Centre, University of Southampton, Southampton SO17 1BJ, UK}
	
\title{On the Role of Longitudinal Currents in Radiating Systems of Charges}
\begin{abstract}
The time derivative of the charge density is linked to the current density by the continuity
equation. However, it features only the longitudinal part of a current density, which is
known to produce no radiation. This fact usually remains unnoticed though it poses a seemingly
serious paradox, suggesting that the temporal variation of a charge density should be also irrelevant
for radiation. We resolve this paradox by showing that the effective longitudinal currents are not
spatially confined even when the time-dependent charge density that generates them is. This
enforces the co-existence of the complementary (i.e. transverse) part of the current density through
the Helmholtz decomposition. We illustrate the mechanics of the underlying non-locality of the
Helmholtz decomposition in the case of a dynamic electric dipole, discussing its practical
implication for underwater antenna communications. More generally, we explain the role of the
Helmholtz decomposition in shaping the structure of the conventional multipole expansion.
\end{abstract}
\maketitle
\section*{Introduction}
In many problems of classical electrodynamics involving electromagnetic radiation it is
methodologically useful to substitute the time-dependent charge density $\rho(t)$, with its
current density equivalent $\ve{j}(t)$, related by the continuity equation (see, for example, \cite{jackson_classical_1999})
\begin{align}
\p_t\rho+\div\ve{j}=0 \ . \label{cont}
\end{align}
However, this generally adopted approach leads to an apparent paradox which, to the best of
our knowledge, has not been discussed (let alone alleviated) in textbooks on electrodynamics.
The tension is caused by the following observation. On the one hand, the time-dependent charge density
gives rise to radiation – accelerating particles and oscillating electric dipoles are the well-known
examples here. On the other hand, the continuity equation is missing (though implicitly) the
transverse part of the current density $\ve{j}_\pp$, which is held responsible for generating radiation fields,
i.e. electric and magnetic fields outside the volume taken by the charge density, see e.g. classic paper \cite{Devaney1973} or textbook \cite{Bo}. Colloquially speaking, longitudinal currents are usually considered to produce no radiation.

Indeed, assuming that the vector field $\ve{j}(r,t)$ is sufficiently smooth and decays
sufficiently fast at infinity, it can be split via the Helmholtz decomposition into the longitudinal
($\rot \ve{j}_\pl = 0$) and transverse ($\div \ve{j}_\pp = 0$) components, and so in the continuity equation $\div \ve{j} = \div (\ve{j}_\pl
+ \ve{j}_\pp) = \div \ve{j}_\pl$. Since longitudinal currents do not radiate, the time-dependent charge density must
also be irrelevant for the radiation.

To resolve this apparent paradox one needs to accept that the longitudinal and transverse parts
of a spatially localized current density are not completely independent, as was briefly noted in \cite{dubovik1974multipole}. Moreover, even when a system of non-stationary charges is confined to
some finite region of space, the transverse and longitudinal components of a current density not
only can exist outside that region but extend to infinity, provided that $\ve{j}_\pl = -\ve{j}_\pp$ there \cite{PhysRev.91.898}. This simple yet, perhaps,
counterintuitive observation is one of the main points that we elaborate on in this paper.

\section{Longitudinal and transverse currents of a localized source \label{sec 1}}
We begin by clarifying how (or rather when) a vector field can be simultaneously transverse and
longitudinal. One usually identifies a vector field with such a property while deriving the Helmholtz
decomposition, but here we are going to avoid the lengthy derivation and offer a brief and perhaps simpler alternative in the style of \cite{griffiths2013introduction}. 

If a vector field $\ve{j}$ (which at the moment is not necessarily a current density) is longitudinal, i.e. $\rot\ve{j}=0$, it can
always be represented as the gradient of some scalar potential $\psi$
\begin{align}
\ve{j}=\grad \psi \ . \label{j from psi}
\end{align}
The requirement for this field to be simultaneously transverse, i.e. $\div \ve{j} = 0$, leads to $\Delta \psi = 0$. If this condition holds everywhere in space the corresponding current is vanishing.\footnote{Assuming that $\ve{j}$ decays sufficiently fast at spatial infinity  it follows $\int d^3\ve{r} \,\,\ve{j}(\ve{r})\cdot \ve{j}(\ve{r})=-\int d^3\ve{r}\,\, \psi \Delta \psi =0$ and hence $\ve{j}(\ve{r})=0$ everywhere.} However, if $\Delta \psi = 0$ holds only outside some region $R$, then $\psi$ can be perfectly non-trivial, and so $\ve{j}$ will be both transverse and longitudinal outside $R$.

In fact, this type of fields is not at all exotic. Electric field of (almost) any static distribution of charges is of this kind. Indeed, consider a static charge density $\rho_0$ confined to $R$. Its scalar potential $\phi_0$ is governed by the Poisson equation\footnote{We introduce subscript 0 to indicate quantities related to this auxiliary static problem.}
\begin{align}
\Delta \phi_0 = -4\pi \rho_0 \ . \label{poisson}
\end{align}

Outside $R$ such a potential naturally satisfies $\Delta \phi_0 = 0$ and, therefore, the electric field generated there $\ve{E}_0=-\grad\phi_0$ is simultaneously longitudinal and transverse. Importantly, this general example also shows that the leakage of the longitudinal field component outside the confined source and up to infinity is almost inevitable.\footnote{There are exceptions though, as discussed in appendix \ref{app nr}.}

Now let us consider a general current $\ve{j}$ with its Helmholtz decomposition into transverse and longitudinal parts $\ve{j}=\ve{j}_\pl+\ve{j}_\pp$. We will show that both $\ve{j}_\pl$ and $\ve{j}_\pp$ generically extend to infinity even when $\ve{j}$ is confined. Scalar potential $\psi$ that defines the longitudinal part $\ve{j}_\pl=\grad \psi$ can be found from condition $\Delta \psi=\div \ve{j}$ or
\begin{align}
\Delta \psi = -\p_t \rho \ , 
\end{align}
where we have used the continuity equation \eqref{cont}. This is mathematically the same problem as \eqref{poisson} with $\psi$ in place of $\phi_0$ and $-\frac{1}{4\pi}\p_t\rho$ in place of $\rho_0$. This means that $\ve{j}_\pl$ is proportional to the time-derivative of the longitudinal part of the electric field produced by the actual charge density
\begin{align}
\ve{j}_\pl=-\frac{1}{4\pi}\p_t \ve{E}_\pl \ .
\end{align}
One could have derived this result directly from the Maxwell equation for the current density
\begin{align}
4\pi\ve{j}=-\p_t\ve{E}+c\rot\ve{H}
\end{align}
by considering its longitudinal component. However, it would then be unclear that $\ve{j}_\pl$ was not simply vanishing. Indeed, one is naturally tempted to conclude that $\ve{E}_\pl= 0$ because the electric field is
transverse in the radiation zone (i.e. $\div \ve{E} = 0$). This conclusion is obviously premature since $\ve{E}_\pl$
and $\ve{E}_\pp$ must be defined everywhere in space (and not only in the radiation zone), and hence $\ve{E}_\pl$ is
(typically) non-zero.

As we have demonstrated above, the longitudinal electric field usually extends beyond the
confines of a static distribution of charges and, thus, the longitudinal component of the current
density $\ve{j}_\pl$ does the same in the dynamic case. The most important consequence of the spatially
delocalized nature of $\ve{j}_\pl$ is that the transverse component of the current density $\ve{j}_\pp$ must also be non-vanishing, even though this cannot be inferred directly from the continuity equation. This simply follows from the fact that the total current density $\ve{j}=\ve{j}_\pl+\ve{j}_\pp$ vanishes outside the source and hence
\begin{align}
\ve{j}_\pp=-\ve{j}_\pl \qquad (\text{outside $R$}) \ . \label{jsum}
\end{align}

Now we are ready to revisit the apparent paradox posed in the introduction. Given that $\ve{j}_\pl$ and $\ve{j}_\pp$ of a localised source are linked, a more precise statement is not that the longitudinal currents are irrelevant to the radiation, but rather that the radiation fields can be expressed solely via the transverse currents. Indeed, a vector potential $\ve{A}$ can be written in the following form (see app. \ref{app Maxwell})
\begin{align}
\ve{A}(\ve{r})=-\f{4\pi}{k^2c}\ve{j}_\parallel(\ve{r})+\f1{c}\int_R d\ve{r}'\,G(\ve{r}-\ve{r}')\ve{j}_\perp(\ve{r'})\label{A via j+j} \ .
\end{align}
Here $\ve{A}$ is non-locally related to $\ve{j}_\pp$, i.e. the distribution of $\ve{j}_\pp$ inside $R$ crucially affects the radiation field outside $R$. However, the longitudinal part of the current density enters $\ve{A}$ only locally. Since Eq. \eqref{A via j+j} holds in any region of space, one can replace $\ve{j}_\pl$ with  $- \ve{j}_\pp$ outside $R$ and, as a result, the
radiation fields will be expressed solely via $\ve{j}_\pp$. 

At the same time, one usually cannot alter $\ve{j}_\pp$ without altering $\ve{j}_\pl$ (or, equivalently, the charge density of the source) and so it would be misleading to state that $\ve{j}_\pl$ is completely irrelevant to the radiation. This is especially true if the radiation occurs in a partially conducting media (such as sea water), where the total current density outside the source does not have to be zero. We consider this case in more detail in the next section.
\section{Electric dipole}
The preceding discussion was rather abstract and we would now like to illustrate it using the simplest non-trivial example of a radiating localized source – a non-stationary electric dipole of negligible size. For the dipole moment $\ve{d}$ placed at the origin $r=0$ the charge and current densities are
\begin{align}
	\rho = -(\ve{d}, \nabla)\,\,\delta(\ve{r}) \label{rho},\qquad \ve{j}= \dot{\ve{d}}\,\delta(\ve{r}) \ .
\end{align}
Neither $\rot \ve{j}$ nor $\div \ve{j}$ vanish everywhere, so $\ve{j}$ contains both longitudinal and transverse parts. Let us explicitly construct $\ve{j}_\parallel$ corresponding to the point dipole. Since $\div \ve{j}_\perp=0$ by definition and $\ve{j}_\parallel=\nabla \phi$, Eq. \eqref{cont} transforms into
\begin{align}
\Delta \phi = (\dot{\ve{d}}, \ve{\nabla})\delta(\ve{r}) \ ,
\end{align}
which has the following solution
\begin{align}
\phi = -(\dot{\ve{d}}, \ve{\nabla})\frac{1}{4\pi r} \ .
\end{align}
The longitudinal part of the corresponding current density is given by the gradient of $\phi$ 
\begin{align}
\ve{j}_\parallel = \grad \phi = -\frac{\dot{\ve{d}}-3(\dot{\ve{d}}, \hat{\ve{r}})\hat{\ve{r}}}{4\pi r^3} \ , \label{jdip}
\end{align}
where $\hat{\ve{r}}=\ve{r}/r$. At any given moment this is nothing but the electric field produced by the static electric dipole with the moment $d_0=-\p_t\ve{d}$ \footnote{Of course for time-dependent $\ve{d}$ the "electric field" in Eq.\eqref{jdip} is time-dependent as well, but at each moment in time it has a shape of the electric field from a static dipole.}. We emphasize that Eq.\eqref{jdip} makes it evident that $\ve{j}_\pl$ extends beyond $r=0$, and, in fact, up to infinity.

If the electric dipole resides in vacuum, $\ve{j}_\parallel$ must be canceled outside $r = 0$ so that the total current density produced externally will remain vanishing. This is possible only by admitting the co-existence of the spatially non-localized complementary, transverse part of the current density of the form 
\begin{align}
	\ve{j}_\pp = \rot^2 \f{\dot{\ve{d}}}{4\pi r} \ .
\end{align}
Explicit computation gives
\begin{align}
\ve{j}_\pp=(\operatorname{grad}\div-\D)\f{\dot{\ve{d}}}{4\pi r}=\nabla(\dot{\ve{d}}, \nabla)\frac1{4\pi r}+\dot{\ve{d}}\,\delta(\ve{r}) \ ,\label{j perp}
\end{align}
where we used identity $\delta(\ve{r})=-\D \frac1{4\pi r}$. It is easy to see that the first term in Eq.\eqref{j perp} is equal to $ -\ve{j}_\pl$, while the second term gives the spatially localized current density of the point electric dipole \eqref{rho}.

\begin{figure}[t] 
	\begin{center}
		\includegraphics[width=0.5\textwidth]{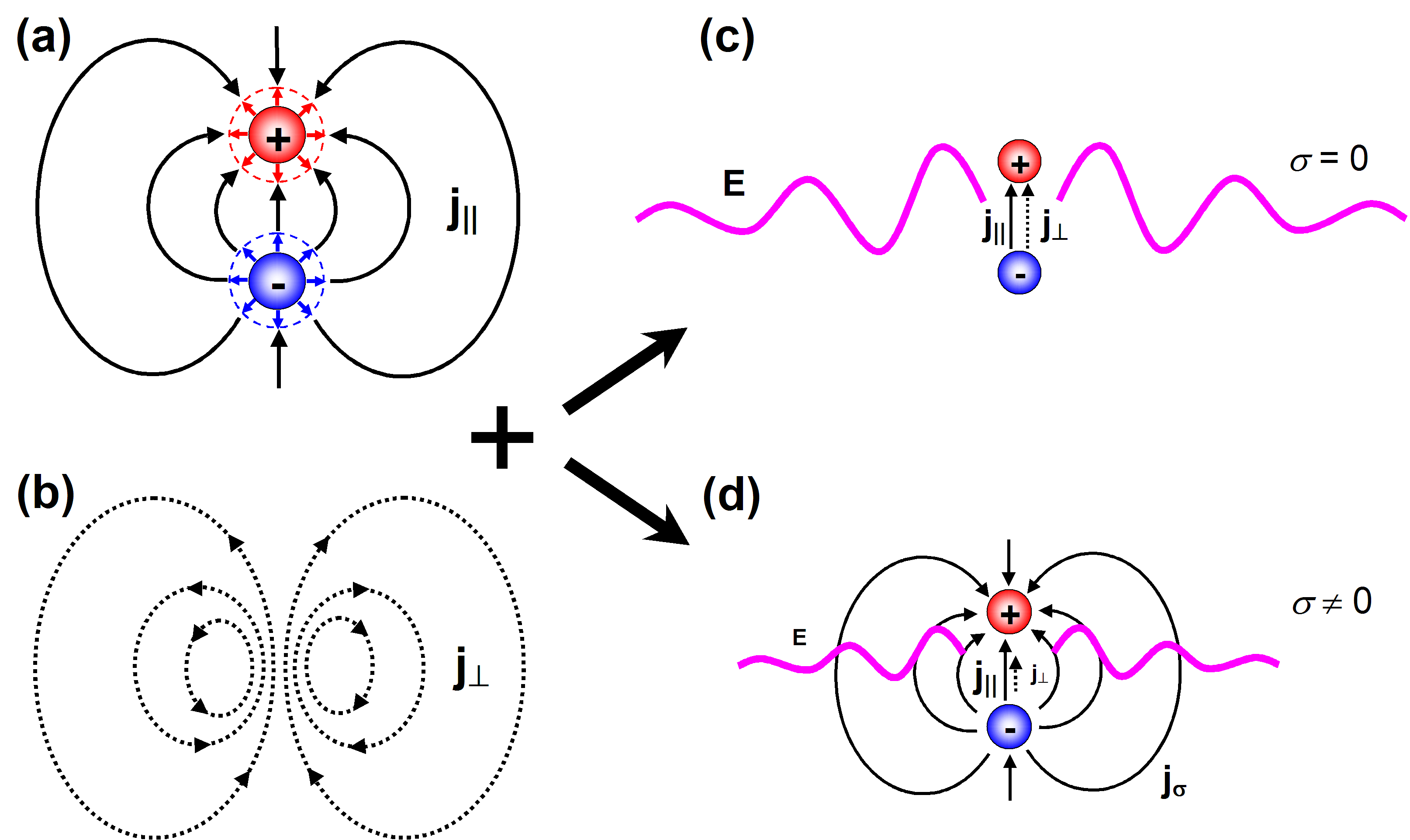}
	\end{center}
\caption{(a) An oscillating electric dipole shown schematically as a pair of non-stationary charge distributions of opposite signs at a moment in time when the charge densities increase. Curved black arrows represent longitudinal currents, which ensure the change in accordance with the continuity equation. (b) The distribution of transverse currents, which can counteract longitudinal currents beyond the confines of the electric dipole. (c) The electric dipole radiates in non-conducting medium, where the cancellation of longitudinal currents by transverse currents must be complete. (d) An electric dipole radiates less efficiently in partially conducting medium since the complete cancellation of longitudinal currents is not required, and so transverse currents (responsible for radiation) are weaker. }
\label{fig1}
\end{figure}

Fig.\ref{fig1} offers a graphical illustration of the above analysis. Without loss of generality we assume that the radiating electric dipole is formed by a pair of spatially localized positive and negative charge distributions (see Fig.\ref{fig1}a). We further assume that the change of the electric dipole moment occurs due to oscillations of the charge densities (rather than the distance between the two). 

We now consider a moment in time when the charge densities increase. According to the continuity equation, an increase in charge density is ensured by currents flowing in (or out) of the volume containing the charges, and those currents are longitudinal. In our case the currents must flow towards the positive charge distribution, where positive charges are accumulated, and away from the negative charge distribution, where positive charges are lost. Considering each charge distribution separately one can already appreciate the fact that the longitudinal currents will formally extend to infinity.

In a physical system longitudinal currents must find their source and drain (due to charge conservation). In our example this means that the currents leaving the negative charge distribution and entering positive charge distribution must be connected everywhere (including infinity), as illustrated in Fig.\ref{fig1}a.
However, in a typical situation – when an electric dipole is located in a dielectric or vacuum – charges cannot be physically transported outside the confines of the dipole and, hence, the longitudinal currents must appear somehow negated.

At this point one has to admit the presence of complementary oscillating transverse currents (see Fig.\ref{fig1}b). Such currents require neither a source nor a drain, and "can be arranged" in space in a way that they will cancel longitudinal currents everywhere apart from the central region, where they have the same density and flow in the same direction as the longitudinal currents. The resulting picture (Fig.\ref{fig1}c) is now intuitively more acceptable, since it shows the electric dipole as a spatially localised charge-current distribution, the radiation of which is allowed through the presence of confined transverse currents.    

One may still wonder why going to the trouble of introducing into the picture spatially non-localized currents for an a priory localized electromagnetic source, especially that those currents cannot be sustained outside the source, and thus appear purely virtual. As we have pointed out earlier, such an approach enables one to show explicitly why a system of non-stationary charges acquires a transverse component of its real current density and, therefore, can radiate. 

Furthermore, and perhaps more importantly, admitting spatially non-localized currents helps one to understand the peculiarities of antenna radiation in partially conducting media, such as sea water or soil \cite{Al-Shammaa2004, Honglei2015}.
Indeed, in such media the longitudinal currents outside the electric dipole in Fig.\ref{fig1}a can be partially sustained (via real conductivity) and, therefore, are not required to be completely negated by transverse currents. Correspondingly, the current balance outside the electric dipole \eqref{jsum} is modified as follows
\begin{align}
\ve{j}_\pp+\ve{j}_\pl=\ve{j}_\sigma \ ,
\end{align}
where $\ve{j}_\sigma$ is the density of the actual longitudinal currents supported by the medium due to its non-vanishing electric conductivity. Since $\ve{j}_\pl$ remains fixed by the continuity equation, $|\ve{j}_\pp| < |\ve{j}_\pl|$ and this inequality becomes only stronger with increasing conductivity of the medium. As a result, not only the emitted electromagnetic waves will decay faster than in vacuum/air (due to absorption) but also the overall radiation efficiency of an electric dipole antenna will become lower even in a well matched case (see Fig.\ref{fig1}d). 

In plain words, the fraction of energy that would have been radiated by the antenna is now circulated in space (and dissipated) via real longitudinal currents. This, in particular, might explain anomalously high attenuation of radiation from electric dipole antennas in sea water and shows why loop (i.e., magnetic dipole) antennas, which do not feature longitudinal currents, are generally better emitters in partially conducting media \cite{Al-Shammaa2004, Honglei2015}.

\section{Implications for the multipole expansion}
Another way to formulate the puzzle that we have started with is to note that in a typical electrodynamic system the charge multipoles, such as an electric dipole or quadrupole, are central to radiation. Their moments, however, are defined solely by the charge density. For instance
\begin{align}
&\ve{D}=\int d\ve{r}\, \ve{r} \rho(\ve{r}) \ ,\\
&Q_{\al\be}=3\int d\ve{r}\, \left(3r_\al r_\be-r^2\delta_{\al\be}\right) \rho(\ve{r}) \ ,
\end{align}
where $ve{D}$ and $Q$ are the electric dipole and quadrupole moments, respectively. The charge density is formally related only to the longitudinal currents $\ve{j}_\pl$ \eqref{cont}  and, therefore, it is seemingly irrelevant for the radiation fields, which can be defined only through transverse currents $\ve{j}_\pp$. Now, of course, we appreciate that the paradox is resolved because $\ve{j}_\pp$ and $\ve{j}_\pl$ are not independent. In particular, the time-derivatives of all electric multipole moments can be expressed via $\ve{j}_\pp$. For instance
\begin{align}
\p_t\ve{D}=\int d\ve{r}\,\, \ve{r} \p_t\rho(\ve{r})=-\int d\ve{r}\,\, \ve{j}=-\int d\ve{r}\,\, \ve{j}_\pp \ .
\end{align}
Here we used the Helmholtz decomposition and the fact that $\int d\ve{r} \ve{j}_\pl=0$, which is not entirely trivial but true, as explained in \ref{volume int}.

It seems however that the non-locality of the Helmholtz decomposition brings about an additional confusion to the structure of the multipole expansion. The full multipole expansion of an arbitrary current density consists of the charge, magnetic and toroidal multipoles. For instance, magnetic and toroidal dipoles are defined by
\begin{align}
\ve{M}=\frac12 \int d\ve{r}\,\, \ve{r}\times \ve{j} \ ,\\
 T_\al=\f{1}{10} \int d\ve{r}\,\, \left(r_\al r_\be-2r^2\delta_{\al\be}\right)j_\be \ .
\end{align}
It is sometimes claimed that the toroidal multipole family represents merely higher-order corrections to the charge multipoles. This is, of course, not correct -- the two families are independent of each other, just as they are independent of the magnetic multipoles, with one notable exception, the leading order.

To explain this we need to recall additional terms of the multipole expansion of a non-stationary charge-current distribution, the mean-square radii \cite{Nemkov2018a}. For any given multipole moment (which can be viewed as a measure of certain angular properties of the charge-current distribution) there is a series of the mean square radii, which characterizes the radial profile the corresponding multipolar mode. For example, the mean-square radii of an electric dipole moment are defined by\footnote{Here and further in this section $\propto$ means equality up to non-zero numerical factors which we do not keep track of.}
\begin{align}
\ve{D}^{(n)}\propto\int d\ve{r}\,\, r^{2n}\ve{r}\rho(\ve{r}) \label{D msr}
\end{align}
and one simply inserts an additional factor of $r^2$ to obtain $\ve{D}^{(n+1)}$ from $\ve{D}^{(n)}$. One can also move in the reverse direction, i.e. from $\ve{D}^{(n+1)}$ to $\ve{D}^{(n)}$, by inserting $\D \rho$ in place of $\rho$:
\begin{multline}
\ve{D}^{(n)}\propto \int d\ve{r}\,\,r^{2n+2}\ve{r}\D \rho(\ve{r})\propto\int d\ve{r}\,\,r^{2n}\ve{r}\rho(\ve{r}) \ .
\end{multline}
If one continuous the reverse transformation further, one will see that the lowest-order term in the series, i.e. the "parent" term (given by $n = 0$) is, of course, the electric dipole moment itself, $\ve{D}^{(0)}=\ve{D}$.

The point we wish to make is that the same procedure applied to the mean-square radii of a toroidal dipole will also yield an electric dipole moment as the lowest-order term of the series (but now corresponding to $n = -1$)
\begin{align}
T^{(-1)}_\al\propto\int d\ve{r}\left(r_\al r_\be-2r^2\delta_{\al\be}\right)\D j_\be\propto D_\al \ ,
\end{align}
since $\D\left(r_\al r_\be-2r^2\delta_{\al\be}\right)=-10\delta_{\al\be}$.
This invites one to conclude that the electric dipole moment  not only is the actual "parent" term of the series of mean-square radii here, but also is the lowest order term of the toroidal multipole family. Note also that for the mean-square radii of any familiar electric or magnetic multipole the above procedure cannot extended beyond $n = 0$, yielding  vanishing result for $n=-1$. Indeed, consider, for example, an electric quadrupole -- the $r$-dependent weight in the integrand satisfies $\D(3r_\al r_\be-r^2\delta_{\al\be})=0$. So despite the fact that the charge and toroidal multipoles are independent families their lowest orders seem to agree.

To show that this coincidence is a direct consequence of the non-locality in the Helmholtz decomposition, as well as to place the above heuristic discussion on a firm ground, we need to define multipole moments precisely. A succinct way to do so is to introduce the following orthonormal basis of vector fields \cite{dubovik1974multipole, Radescu2002}
\begin{align}
&\ve{\c{F}}_{lmk}^{(-)}(\ve{r})=\f{\nabla \c{F}_{lmk}(\ve{r})}{ik}\nonumber\ ,\\
&\ve{\c{F}}_{lmk}^{(0)}(\ve{r})=\f{\rot \left(\ve{r}\c{F}_{lmk}(\ve{r})\right)}{-i\s{l(l+1)}}\nonumber\ ,\\
&\ve{\c{F}}_{lmk}^{(+)}(\ve{r})=\f{\rot^2 \left(\ve{r}\c{F}_{lmk}(\ve{r})\right)}{-ik\s{l(l+1)}} \ . \label{f def}
\end{align}
where $\c{F}_{lmk}=j_l(kr)Y_{lm}(\ve{n})$, $j_l$ is the Bessel function and $Y_{lm}$ are the spherical harmonics. Their origin and properties are explained in appendix \ref{app functions}. For our purposes it is most important to note that the so-called form-factors \cite{dubovik1974multipole, dubovik1975multipole, Dubovik1990, Gongora2006} 
\begin{align}
m^{(\lambda)}_{lm}(k)=\int d\ve{r}\, \ve{j}(\ve{r}) \left(\ve{\c{F}}^{(\lambda)}_{lmk}(\ve{r})\right)^*
\end{align}
directly encode both the multiple moments and mean-square radii a current distribution. More specifically, form-factor $m^{(-)}_{lm}(k)$ corresponds to the charge multipoles\footnote{This is clear for example from the fact that $\ve{\c{F}}^{(0)}$ and $\ve{\c{F}}^{(+)}$ are orthogonal to $\ve{j}_\pl$.}, $m^{(0)}_{lm}(k)$ -- to magnetic multipoles and $m^{(+)}_{lm}(k)$ -- to toroidal multipoles. Indices $lm$ in each form-factor describe the multipole being probed (dipole, quadrupole etc.), while $k$ defines the radial profile of each multipolar mode and corresponds to the mean-square radii. 

The multipoles themselves are proportional to form-factors at $k=0$. For instance (up to the usual transformation between the Cartesian and spherical basis)
\begin{align}
\ve{D}\propto m^{(-)}_{1,i}(k=0) \ .
\end{align}
The mean square radii appear as coefficients in the Taylor expansion of $m^{(\lambda)}_{lm}(k)$. For example, the mean-square radii of the dipole moment \eqref{D msr} are proportional to 
\begin{align}
D^{(n)}\propto \frac{d^{2n}}{dk^{2n}}m^{(-)}_{1, i}(k)\Big|_{k=0} \ .
\end{align}
Now, form-factors $m^{(\lambda)}_{lm}(k)$ are in general independent functions and describe three independent multipole families. However it turns out that functions $\ve{\c{F}}_{lmk}^{(-)}$ and $\ve{\c{F}}_{lmk}^{(+)}$  have the same behavior at small $k$
\begin{align}
\ve{\c{F}}_{lmk}^{(+)}\approx \s{l(l+1)}\ve{\c{F}}_{lmk}^{(-)},\qquad k\to0 \ .\label{+-equiv}
\end{align}
As a consequence, the multipole moments defined by $m^{(-)}_{lm}(0)$ and $m^{(+)}_{lm}(0)$ coincide (up to numerical factors).  Crucially though, $m^{(-)}_{lm}(k)$ and $m^{(+)}_{lm}(k)$ cease to agree beyond $k=0$ and, hence, the charge and toroidal mean-square radii are in general all different. 

The above analysis formalizes our heuristic derivation at the beginning of this section, which showed that the toroidal dipole moment has the usual electric dipole moment as its "parent multipole" and that the same applies to all toroidal multiples. Because of this coincidence, the toroidal family is usually defined to start one order higher, i.e. the 1st mean square radii of form-factors $m^{(+)}_{lm}(k)$ are considered to be the primary toroidal multipoles \cite{dubovik1974multipole}. In retrospect, it would be, perhaps, more consistent to refer to the quantities $m^{(+)}_{lm}(k=0)$ (and thus $\propto m^{(-)}_{lm}(k=0)$) as toroidal rather than charge multipole moments, because the mean-square radii of the toroidal multipoles all contribute to radiation, while the charge mean-square radii do not \cite{Nemkov2018a}.

Finally, let us emphasize that the noted relation between the charge and toroidal multipoles is conditioned by two factors. The first is the apparent numerical coincidence\footnote{Form-factor functions \eqref{f def} are in fact uniquely fixed by Helmholtz decomposition together with requirements of irreducibility and parity \eqref{parity}, so this apparent numerical coincidence has deeper roots. } stated in Eq.\eqref{+-equiv}. The second is that although charge multipole moments are originally defined only via the longitudinal currents, while toroidal moments only via transverse currents, the non-locality of the Helmholtz decomposition imposes a connection between them. In particular, note that $k\to0$ limit (when the charge and toroidal form-factors agree) corresponds to the infinite-wavelength limit. In this regime the distinction between $\ve{j}_\pl$ and $\ve{j}_\pp$ becomes negligible, as they only differ in a finite region, which can not be resolved by a wave with an infinitely large  wavelength.
\section*{Acknowledgments}
We thank Alexey Basharin for useful discussions. The work of NN is partly supported by Leading Research Center on Quantum Computing (Agreement no.  014/20).
\appendix
\section{Maxwell's equations \label{app Maxwell}}
Here we collect our conventions regarding Maxwell's equations. We work with harmonic time-dependence, frequency is denoted by $\omega$, $k=\omega/c$ is the wavenumber. Maxwell's equations can be written as
\begin{eqnarray}
\ve{j}(\ve{r})=\f{c}{4\pi}\left(-(k^2+\Delta)\ve{A}(\ve{r})+\grad\div \ve{A}(\ve{r})\right)\ , \label{current via vector potential}
\end{eqnarray} 
where $\ve{A}(\ve{r})$ is a vector potential in the Weyl gauge (vanishing scalar potential). In this gauge the electric and magnetic fields are related to the vector potential as follows
\begin{align}
\ve{E}(\ve{r})=ik \ve{A}(\ve{r}),\qquad \ve{H}(\ve{r})=\rot \ve{A}(\ve{r}) \ .
\end{align}
Conversely, assuming that the vector potential decays at infinity it can be found by solving \eqref{current via vector potential}
\begin{align}
\ve{A}(\ve{r})=\f1{k^2c}\int d\ve{r}'\,G(\ve{r}-\ve{r}')\left(k^2\ve{j}(\ve{r}')+\grad\div \ve{j}(\ve{r'})\right) \ ,
\end{align}
where $G(\ve{r}-\ve{r}')=\f{e^{ik|\ve{r}-\ve{r}'|}}{|\ve{r}-\ve{r}'|}$ is the Green function for Helmholtz equation $(\D+k^2)G(\ve{r})=-4\pi \delta^{(3)}(\ve{r})$.

The transverse part of the current satisfies $\div \ve{j}_\perp=0$ and gives rise to the transverse part of the vector potential
\begin{align}
\ve{A}_\perp(\ve{r})=\f1{c}\int d\ve{r}'\,G(\ve{r}-\ve{r}')\ve{j}_\perp(\ve{r'}) \ .
\end{align}
The longitudinal part of the current satisfies $\grad \div \ve{j}_\parallel=\D \ve{j}_\parallel$ and, hence 
\begin{align}
\ve{A}_\parallel(\ve{r})=-\f{4\pi}{k^2c}\ve{j}_\parallel(\ve{r}) \ .
\end{align}
Outside of a localized source $\ve{j}_\perp=-\ve{j}_\parallel=0$ and, therefore, both $\ve{A}_\perp$ and $\ve{A}_\parallel$ can be expressed via $\ve{j}_\perp$ alone.
\section{Non-radiating charge densities \label{app nr}}
In the main text we have been careful to make reservations while claiming that any charge density will be consequential for the radiation fields. There is a notable exception. Let us reconsider the Poisson equation \eqref{poisson} and ask whether there exist a charge density $\rho_0$ which actually does not produce a potential $\phi_0$ (and hence the electric field) outside its domain of definition. The answer is yes, and it is simple to describe all such densities. Assume that $\rho_0$ is a Laplacian of some function $\phi$, i.e. $\rho_0=\D \phi$ and $\phi$ is zero outside $R$. Then from \eqref{poisson} it follows that the potential $\phi_0$ is just equal to $\phi$ up to constants and hence is itself confined (the argument is almost tautological). 

This implies the existence of charge densities that do not produce any electric fields outside. Translated into the language of the Helmholtz decomposition this means that some modifications of the longitudinal current do not affect the transverse part. Namely, one can add to $\ve{j}$ a term of the form $\grad\phi$ with any confined function $\phi$.

\section{Volume integral of longitudinal currents}
Here we show that for a spatially confined current $\ve{j}$ its longitudinal part $\ve{j}_\pl$ satisfies
\begin{align}
\int d\ve{r}\, \ve{j}_\pl=0 \ , \label{volume int}
\end{align}
which makes it possible to express all multipole terms as integrals of $\ve{j}_\pp$ alone. As explained in Sec. \ref{sec 1} relation between $\ve{j}$ and $\ve{j}_\pl$ arises through a Poisson equation
\begin{align}
\Delta \psi(\ve{r}) = \div \ve{j}(\ve{r}),\qquad \ve{j}_\pl(\ve{r})=\grad \psi  \ ,
\end{align}
which is solved by
\begin{align}
\psi(\ve{r})=-4\pi\int d\ve{r}'\, \frac{\div \ve{j}(\ve{r}')}{|\ve{r}-\ve{r}'|} \ .
\end{align}
Because the source is spatially confined the range of $\ve{r}'$ is bounded and for large enough $\ve{r}$ we can write
\begin{multline}
\psi(\ve{r})=-\frac{4\pi}{r}\int d\ve{r}' \div \ve{j}(\ve{r}')+\\\f{4\pi}{r^2}\int d\ve{r}' (\ve{n}, \ve{r'})\div \ve{j}(\ve{r}')+O\left(\frac{1}{r^3}\right)
,\end{multline}
where $\ve{n}=\ve{r}/r$. The first term here is zero by the Gauss theorem, while the second can be rewritten
\begin{align}
\psi(\ve{r})=\f{4\pi}{r^2} (\ve{n}, \ve{V})+O\left(\frac{1}{r^3}\right),\quad \ve{V}=\int d\ve{r}'\ve{r'}\div \ve{j}(\ve{r}') \ .
\end{align}
Here $\ve{V}$ is an analog of the dipole moment but for our purposes this interpretation is not important. $\ve{V}$ is just a constant vector characterizing current distribution $\ve{j}$.

Now we are ready to prove \eqref{volume int}. Consider
\begin{multline}
\int d\ve{r}\, \ve{j}_\pl = \int d\ve{r}\, \grad \psi=\\\lim_{R\to\infty}\int_{r=R}  4\pi r^2 d\ve{n}  \left(\frac{(\ve{n},\ve{V})}{r^2}+O\left(\frac{1}{r^3}\right)\right) \ .
\end{multline}
The leading term vanishes due to identity $\int d\ve{n}\, \ve{n}=0$ (the average of the normal vector over a unit sphere is zero), while the subleading terms vanish in the $R\to\infty$ limit. This establishes \eqref{volume int}.
\section{Multipole form-factors \label{app functions}}
Functions $\ve{\c{F}}_{lmk}^{(\lambda)}$ introduced in \eqref{f def} are regular solutions to the Helmholtz equation
\begin{align}
(\Delta+k^2)\ve{\c{F}}_{lmk}^{(\lambda)}=0
\end{align}
and satisfy orthogonality and completeness relations
\begin{align}
&\int d\ve{r}\, \ve{\c{F}}^{(\lambda)}_{lmk}(\ve{r})\cdot \ve{\c{F}}^{*(\lambda')}_{l'm'k'}(\ve{r})=\f{(2\pi)^3}{k^2}\delta^{\lambda\lambda'}\delta_{ll'}\delta_{mm'}\delta(k-k')\nonumber\ ,\\
&\sum_{lmk\lambda} \left[\ve{\c{F}}^{(\lambda)}_{lmk}(\ve{r})\right]_i\left[ \ve{\c{F}}^{*(\lambda)}_{lmk}(\ve{r}')\right]_j =(2\pi)^3\delta_{ij}\delta(\ve{r}-\ve{r}') \ .
\end{align}
Under parity transformations $\ve{x}\to -\ve{x}$ they behave as
\begin{align}
\c{F}_{lmk}^{\lambda}(-\ve{x})=(-)^{l+\lambda}\c{F}_{lmk}^{\lambda}(\ve{x}) \label{parity} \ .
\end{align}
The scalar functions $\c{F}_{lmk}=j_l(kr)Y_{lm}(\ve{n})$ that are used to define $\ve{\c{F}}^{(\lambda)}_{lmk}$ also solve the Helmholtz equation and have the following $k\to0$ asymptotic
\begin{align}
\c{F}_{lmk}(\ve{r})=\frac{4\pi i^l}{(2l+1)!!} (k r)^l Y_{lm}(\ve{n})(1+O(k)) \ ,
\end{align}
from which \eqref{+-equiv} can be derived.

\bibliography{bibfile,/home/idnm/Dropbox/hep/Sheets/library.bib}
\end{document}